\documentclass{elsart}

\usepackage{graphicx}

\usepackage{amssymb}

\begin{document}

\begin{frontmatter}



\title{Self-organized network design by \\
link survivals and shortcuts}


\author[JAIST]{Yukio Hayashi}, 
\author[JAIST]{Yuki Meguro}

\address[JAIST]{Japan Advanced Institute of Science and Technology, 
Ishikawa 923-1292, Japan}

\begin{abstract}
One of the challenges for future infrastructures
is how to design a network with high efficiency and
strong connectivity at low cost.
We propose self-organized geographical networks
beyond the vulnerable
scale-free structure found in many real systems.
The networks with spatially concentrated nodes
emerge through 
link survival and path reinforcement on routing flows
in a wireless environment with a constant transmission range of a node.
In particular, we show that
adding some shortcuts induces both the small-world effect
and a significant improvement of the robustness
to the same level as in the optimal bimodal networks.
Such a simple universal mechanism
will open prospective ways for several applications
in wide-area ad hoc networks, smart grids, and urban planning.
\end{abstract}

\end{frontmatter}

\section{Introduction}
The self-organization of a network appears 
in a variety of subjects: 
biological growth, river formation, urban planning, 
and technological infrastructure design.
They are too diverse to comprehend the complex processes
involved in chemical, physical, and socioeconomic 
phenomena.
While we do not insist on the detailed processes, 
we may find a common mechanism for generating 
efficient and robust networks at low cost
in an optimization principle.
In particular, a point of contact in the fields of biology, 
computer science, and physics is important for understanding 
a foundation for the self-organization mechanism.

Some biologically-motivated algorithms may be useful, e.g., 
the slime mold {\it Physarum polycephalum} 
forms efficient, economic, and 
fault-tolerant networks similar to the railway system in the 
Tokyo metropolitan area \cite{Tero10}. 
In the networks, each node tends to connect to neighbor nodes 
in order to approach food sources, 
and some links which do not contribute to energy transport 
gradually disappear.
Thus, short paths between 
any two nodes are self-organized on a planar space.
A similar mechanism based on diffusive growth following chemotaxis 
is applied in the 
modeling of leaf venation patterns \cite{Runions05}
and morphogenesis \cite{Maree01,Maree02}.
Moreover, in human trail systems \cite{Helbing97}, 
frequently used trails are more strongly marked, 
but rarely used trails
are destroyed by the weathering effect.
Apart from detailed processes, 
the key concepts of self-organization are
the selective reinforcement of preferred routes and 
the removal of redundant connections.
Removing the weakest links is also economically reasonable 
in the natural selection for 
maintaining efficient transportation and communication 
networks \cite{Xie08}.

In complex network science, 
several models have been proposed for the 
coevolution of network formation and opinion spreading with the 
phase transition of the community size \cite{Holme06,Zanette06} 
and for the coupled dynamics with network evolution and packet flows 
by random walkers \cite{Noh08,Noh09}; 
the hub emergence is parametrically controllable in 
the phase transition of the topological structure. 
Another idea of path reinforcement is introduced 
by iteratively adding a bypass between 
nodes separated by two hops on frequently traveled paths
\cite{Ikeda07,Ikeda08}. 
Through the induced diffusion process by random walkers,
quasi-complete and scale-free (SF) networks emerge from the 
initial 1D-chain and 2D-lattice, respectively.
However, in the topological formations 
involved with diffusive flow dynamics
\cite{Holme06,Zanette06,Noh08,Noh09,Ikeda07,Ikeda08}, 
the geographical positions of nodes are ignored or restricted in 
spite of their importance at least 
for transportation and communication networks.
In addition, 
these network structures are not necessarily desirable 
due to the large cost of maintaining many links and 
the vulnerability against hub attacks \cite{Albert00}.

On the other hand, 
a planar network emerges through directional growth 
just like in leaf venation patterns \cite{Runions05}
for connecting to the geographical neighbors 
from new nodes whose positions follow a given spatial distribution 
related to the 
population density in city road formation 
\cite{Barthelemy08}. 
It is also an attractive feature 
that in reaction-diffusion networks \cite{Xuan10}, 
various topologies, 
from homogeneous to heterogeneous structures, 
are controlled by rewirings 
according to the change of an external condition. 
In the typical model for ad hoc communication \cite{Rajan08} 
related to adaptive linking mechanisms, 
a clustered network emerges on the random positions of nodes
connected within a transmission range, 
in order to prevent energy consumption.
Thus, beyond an extremely heterogeneous structure like SF networks, 
challenges to the self-organization of
both efficient and robust networks are continued successfully. 

In this paper, we aim to develop a new design method for the 
self-organization of networks in a wireless environment. 
The issues of connectivity, scalability, 
routing and topology control in
ad-hoc networks have been pointed out with 
the importance of mathematical models motivated by 
graph theory \cite{Rajan08}.
We further consider the strong effects of shortcuts 
on communication efficiency and the robustness of
connectivity.

\section{Network models} \label{sec2}
\subsection{Initial configuration}
Let us consider an initial network as 
a well-studied model for wireless ad hoc communication,
which is called a unit disk graph (UDG) \cite{Rajan08} 
in computer science.
The decentralized complex system design 
is very important in challenging discussions for 
future network infrastructures \cite{Krause06}.
Each node corresponds to a base station for multi-hop communication in a
wireless environment. 
Without any biased condition, 
the positions of nodes are distributed uniformly at random
in the normalized $1.0 \times 1.0$ area.
We assume that each node has a single transmission range 
denoted by $A$. 
In the disk area with the radius $A / \sqrt{N_{0}}$ centered at a node, 
the node can directly communicate with 
its geographic neighboring nodes.
Here, $N_{0}$ is the network size (the total number of nodes)
of a UDG.
In other words, this condition for proximity connections
is equivalent to $d_{ij} < A / \sqrt{N_{0}}$, 
where $d_{ij}$ denotes the  Euclidean distance 
between two nodes $i$ and $j$ on a plane, 
and $1 / \sqrt{N_{0}}$ is the normalization factor.
Note that proximity connections are economically appropriate 
for transportation and communication networks,
while low consumption for the transmission 
involves a weak connectivity.
One of the issues related to the UDG is 
how large the transmission range needs to be connective between 
any nodes via a
path of multi-hops in the networks.

\begin{figure}[htb]
  \begin{minipage}[htb]{.47\hsize}
   \includegraphics[height=85mm,angle=-90]{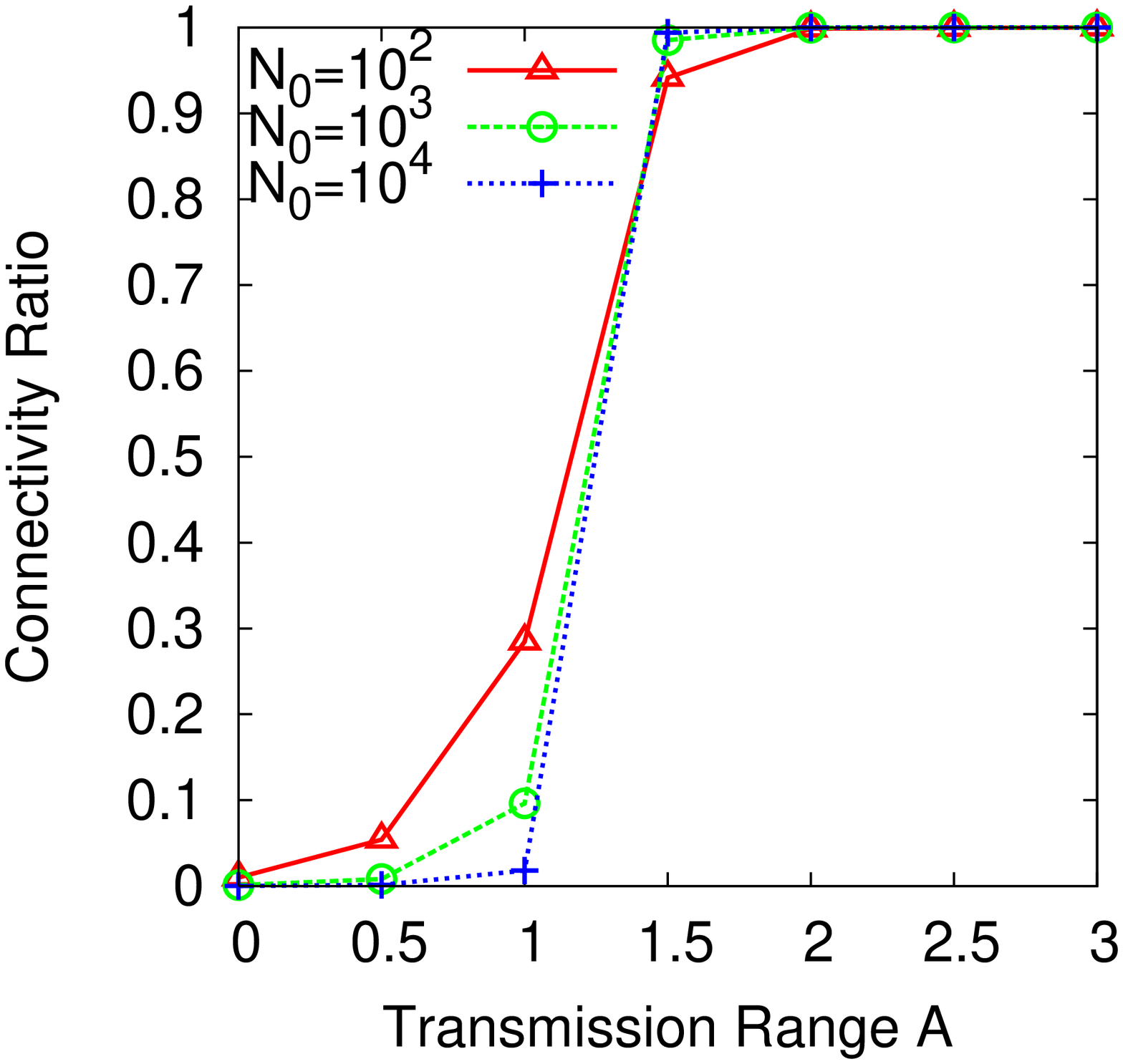} 
    \begin{center} (a) \end{center}
  \end{minipage}
  \hfill
  \begin{minipage}[htb]{.47\hsize}
   \includegraphics[height=85mm,angle=-90]{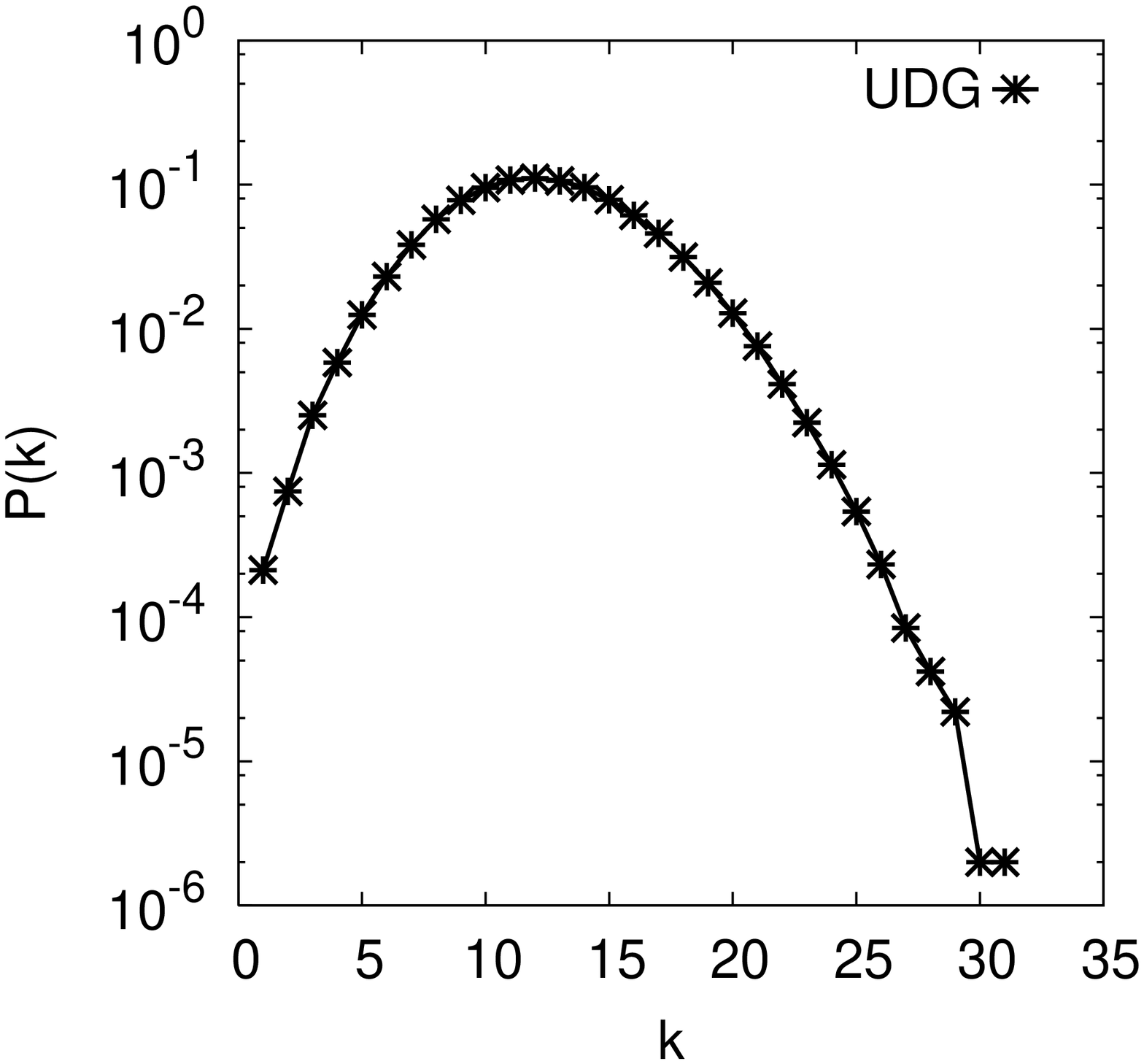} 
    \begin{center} (b) \end{center}
  \end{minipage}
\caption{(Color online) Basic property of the UDG. 
(a) Connectivity ratio vs. transmission range $A$. 
(b) Degree distribution in the UDG for $A = 2.0$ and $N_{0} = 10000$.
Each value is obtained by the average of $50$ realizations.}
\label{fig_prop_udg}
\end{figure}

Figure \ref{fig_prop_udg}(a) shows the phase transition of the
connectivity for a transmission range $A$.
This result is consistent with that 
presented in Refs. \cite{Dall02,Onat08}.
We define the connectivity ratio by a relative size $S / N_{0}$, 
$S$ denotes the number of nodes which can reach each other 
via direct or multi-hop paths in the largest connected component (also
called the giant component). 
In this paper, we use $A = 2.0$ to be connective
via a path between two nodes in almost all pairs. 
Although denser connections are obtained at each node 
for a larger value of $A$, it is numerically confirmed that 
our discussion is not affected 
by the results after this section.
Because redundant links are removed in the course of time, 
even if there are many initial connections,
then the remaining (not removed but surviving) links are 
almost the same as long as there is an initial high connectivity ratio. 
As shown in Fig. \ref{fig_prop_udg} (b), 
a UDG for $A = 2.0$ consists of small degrees in the distribution with 
an exponential decay.
Therefore, it has no hub, since the maximum degree is small as 
$k_{max} \sim O(\log N_{0})$ estimated from 
$\int_{k_{max}}^{\infty} P(k) \sim 1/N_{0}$, 
$P(k) \sim e^{- \gamma k}$, and $\gamma > 0$.

\subsection{Communication request}
At the beginning, 
to simplify the discussion, we consider only two types of nodes: 
the normal node and the active node corresponding to the food source
\cite{Tero10}.
We assume that the type is initially fixed at each node, 
for example, we set the number of active nodes are at $10 \%$ of the
total $N_{0}$, and the remaining $90 \%$ are normal nodes.
The positions of nodes are given by the UDG.

At each time step $t$, $R = 0.1 N_{t}$ packets are generated 
over the network, where $N_{t}$ is the number of surviving nodes 
at time $t$, as mentioned in Subsection 2.4.
Some normal and active nodes are removed from the initial UDG in the
self-organization of the network.
Here, a communication request stochastically occurs between two
nodes of the source and the terminal 
(where it is also called a destination in computer science) as follows: 
a packet is generated at rates $p_{n}:p_{a} = 1:1000$
from normal and active source nodes. 
The terminal nodes are also randomly chosen at the rates.
In other words, an active node is $1000$ times more likely to be 
chosen as the source or terminal than a normal node\footnote{From 
$0.1 N_{t} p_{a} + 0.9 N_{t} p_{n} = R$ and $p_{a} = 1000 p_{n}$,
we derive the packet generation probabilities 
$p_{n} = 1/1009$ and $p_{a} = 1000/1009$ at a node per step.}.
Note that 
there is a one-to-one correspondence between source/terminal and packet.
Thus, for $R$ packets per step, 
the source and the terminal are randomly chosen for normal and active
nodes according to the rates.

\begin{figure}[htb]
  \begin{center}
   \includegraphics[height=95mm,angle=-90]{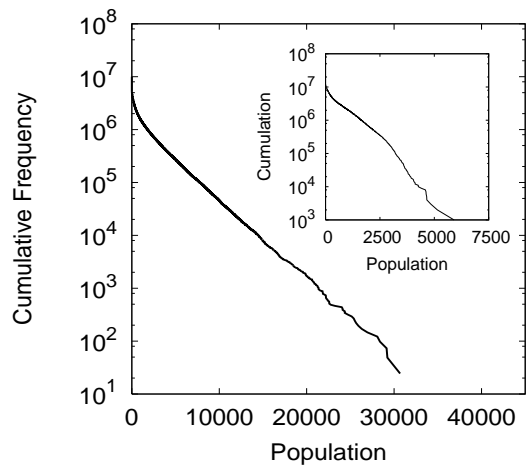} 
  \end{center}
\caption{Cumulative frequency of population in the territory of a node
in UDGs over 50 realizations for $N_{0} = 1000$. 
Inset shows the cumulation of population data in 
the original $160 \times 160$ mesh blocks in $80 km^{2}$
(see Fig. \ref{fig_visualize} for the visualization). 
Both curves are nearly straight lines characterized as 
exponential distributions.}
\label{fig_cum_pop}
\end{figure}

We can extend this simple setting of $p_{n}$, $p_{a}$, 
and the fraction of nodes to any probability distribution 
for choosing the source and the terminal 
in a more realistic situation; 
packets are generated proportionally to a population in the territory
of a node.
By using a statistical population data on a geographical map, 
each mesh block (corresponding to a set of users) is assigned to 
the nearest access node (corresponding to a base station) 
in the sense of the Euclidean distance.
The merged region by the assigned blocks to a node is its 
territory. 
Then, the population in the territory 
is defined by a sum of data at the assigned blocks. 
Figure \ref{fig_cum_pop} shows the cumulative population of nodes
whose positions are uniformly at random for the initial setting of 
UDGs on a geographical map. 
Note that the density function of the population is nearly exponential, 
since the cumulative distribution is almost linear in a semi-log plot, 
and the derivative of an exponential function is also exponential. 
We confirm similar exponential-like distributions 
for several population data at 
other locations measured by the Japan Statistical Association. 
In such real data, it is common 
that big cities are located apart from each other
while some nodes with high populations in their territories 
concentrate in and around a city. 
Thus, the selection of a node as the source or the terminal of a packet 
is proportional to the population in its territory under 
the normalization to be a probability distribution. 
For $R$ packets per step, this process is repeated. 
We call it the packet generation according to the population.
Although probabilistic selections of the source and the terminal 
are due to a general problem setting for the simulation, 
in practice, their nodes should be chosen 
by an objective determined beforehand. 
Therefore, 
it is natural for a user to know the position of a terminal 
without global information.
This explanation supports 
a decentralized routing with only local information 
in the next subsection.

\subsection{Routing protocol}
\begin{figure}[htb]
\begin{center}
 \includegraphics[height=100mm]{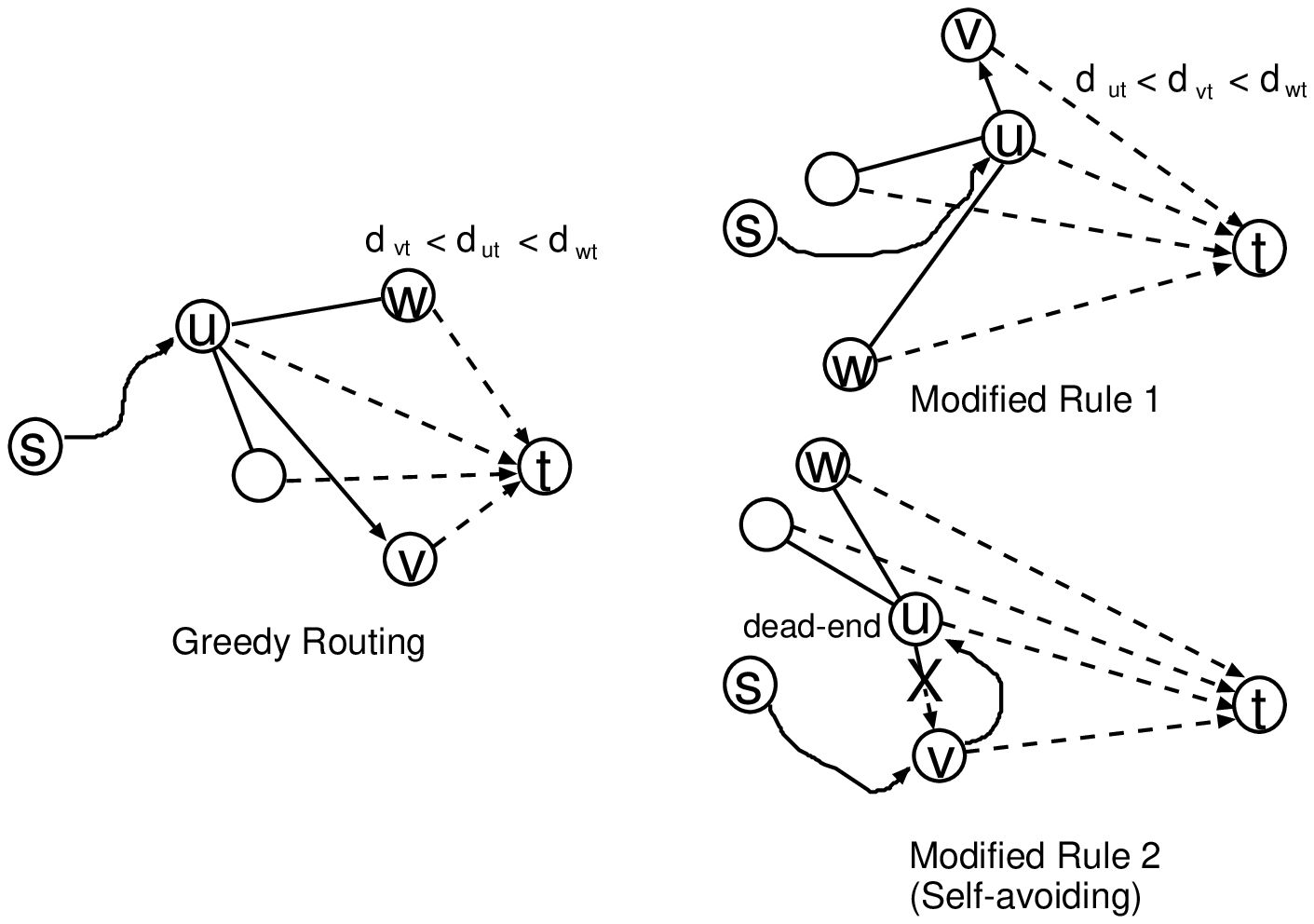} 
\end{center}
\caption{Modified greedy routing. 
The solid straight line denotes a direct link to $v$ or $w$
in the set 
${\mathcal N}_{u}$ of connecting one-hop neighbors, 
and the dashed line denotes $d_{vt}$ 
measured by the Euclidean distance from $v \in {\mathcal N}_{u}$
to the terminal $t$. 
The curved line represents a path from the source $s$ to 
the current resident node $u$ of a packet.}
\label{fig_modified_greedy}
\end{figure}

Instead of the random walk \cite{Ikeda07,Ikeda08}, 
we consider a greedy routing \cite{Fan06} developed in computer science,
as shown on the left of Fig. \ref{fig_modified_greedy}.
It is also called a greedy searching or greedy forwarding algorithm.
In order to forward a packet from the resident node $u$
to a closer position to its terminal $t$
in the sense of the Euclidean distance on a plane, 
a node $v$ is chosen in the set ${\mathcal N}_{u}$ of connecting
one-hop neighbors of $u$, 
if the distance $d_{vt}$ between nodes $v$ and $t$ is minimum:
$d_{vt} \leq d_{wt}$, $v \neq w$, 
$\forall w \in {\mathcal N}_{u}$,
where ties for the minimum are broken randomly in ${\mathcal N}_{u}$.
In this paper, the protocol is modified 
to that a packet is always transferred to a connecting neighbor 
$v \in {\mathcal N}_{u}$ even if $d_{vt} > d_{ut}$ 
(see the top right of Fig. \ref{fig_modified_greedy}).
Moreover, to avoid falling into a trap of cycles, 
we apply the self-avoiding rule: a packet is not transferred to 
nodes that have already been visited 
(see the bottom right of Fig. \ref{fig_modified_greedy}), 
whose history is stored in memory. 
This deterministic routing can act with only local information 
(e.g. positions of one-hop neighbors measured by a GPS)
even for any topological change including the mobility of nodes.
Note that all existing packets 
independently move to one-hop neighbors by a one-time step.
If a packet either arrives at its terminal 
or encounters a dead-end, it is removed.

\subsection{Network self-organization}
We propose the following efficient networks that are 
self-organized through two phases: 
the removal of links and the addition of shortcuts 
depending on traffic flows.
Here, 
the simultaneous implementation of the two phases is outside the scope of
this fundamental proposal, since the network generated by the
simultaneous implementation destroys the initial structure of a UDG in a
wireless environment 
and does not stably converge to a topological structure 
except with the trivial connections between active nodes.
How to simultaneously implement them will be an issue.

First, we explain the base model for self-organization.
In a {\bf Link Survival(LS)} network, 
if a packet passes a link on the modified greedy routing, 
the weight is increased by $1$. 
After the forwarding process for all packets, we check whether
all link weights 
are decreased by $1$ with probability $p_{d} = 0.1$. 
When the weight reaches $0$ after the decreases chosen with $p_{d}$
through several steps, 
then the link is removed, 
and isolated nodes without any links are also removed. 
From an initial UDG with size $N_{0}$, 
both the increase and the decrease of link weights 
are repeated until $T = 3 \times 10^{4}$ steps in the steady state. 
The initial weights are set to 5 for all links
in order not to be removed before packets reach them.

Next, we consider a {\bf Path Reinforcement(PR)} network
extended from the LS network by adding shortcuts. 
After stopping the removal of links at $T$, 
we continue the network generation and the transfer of packets to make
$10$ $\%$ or $30 \%$ of shortcuts 
for the total number of surviving links in the LS network. 
Each shortcut link is added between the current resident node 
of a randomly chosen packet and a randomly chosen node from the nodes 
that have already been visited on the
routing path before arriving at its terminal. 
Such a direct shortcut between nodes separated by more than two hops
suppresses the forming of a quasi-complete graph \cite{Ikeda07,Ikeda08}
in the hop-by-hop connections. 
To compare the communication efficiency and the robustness of
connectivity with LS and PR networks, 
we further consider a {\bf Random Shortcut(RS)} network. 
It is constructed by adding $10$ $\%$ or $30 \%$ of shortcuts between
two randomly chosen nodes independently from the positions of packets on
the LS network.
We emphasize that, in the previous studies for SF \cite{Hayashi07} and
multi-scale quartered (MSQ) \cite{Hayashi09} networks, 
the robustness is significantly improved by adding 
such random shortcuts. 

We do not discuss how to implement shortcut links in practice,
since the problem will be solved by future wireless technologies, 
such as control of directivity and beam power, 
a hybrid network with wired links, etc.
In addition, the values of $p_{d}$ and $T$ are set to be 
merely convenient for the simulation. 
For other values, 
we obtain results that are similar to those presented 
in the next section, 
provided that the value of $p_{d}$ 
is not too large to maintain the connectivity.
Although almost all surviving links do not change any longer 
around $1000 \sim 2000$
steps, we take sufficient time for $T$.

\begin{figure}[htb]
  \begin{center}
    \includegraphics[height=137mm]{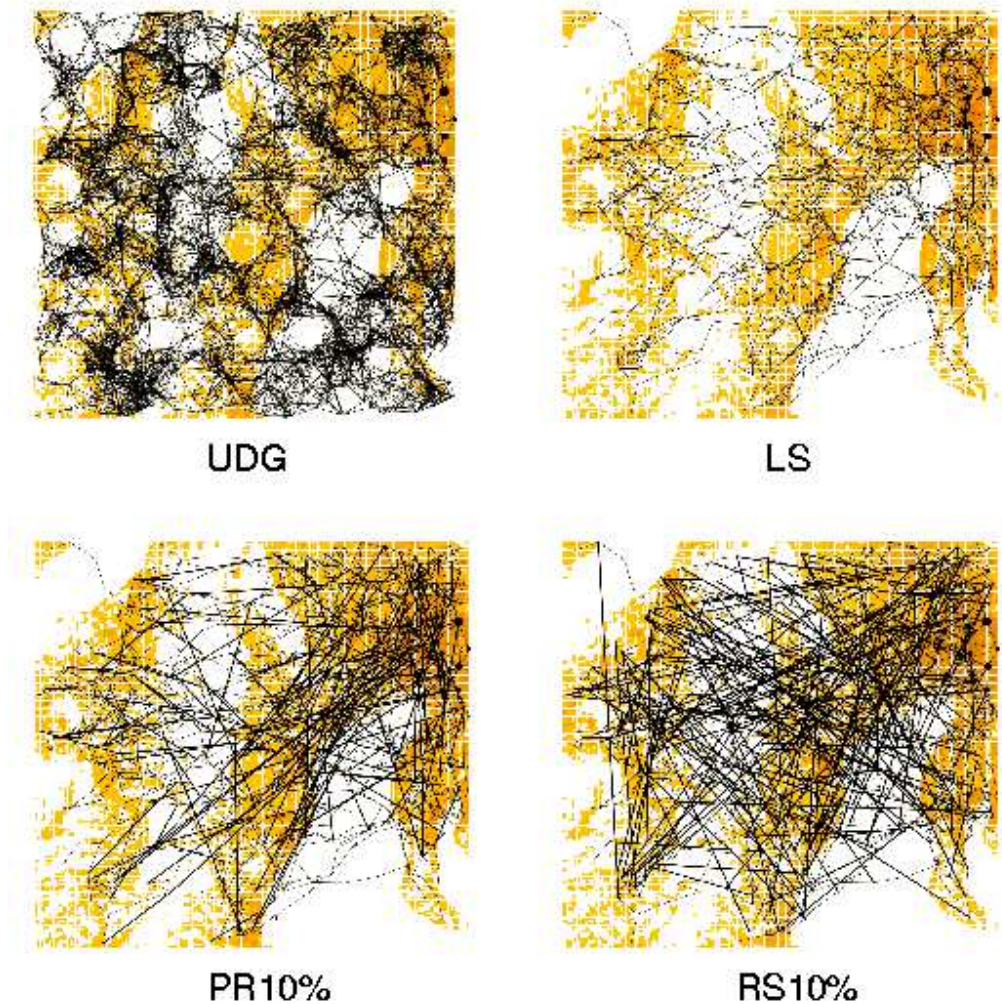} 
  \end{center}
\caption{(Color online) Visualization of the networks 
generated from a UDG with the initial size $N_{0} = 1000$.
In the LS, PR, and RS networks, 
the surviving nodes concentrate in high-population areas 
for which data is provided by the Japan Statistical Association.
The gradation of 
mesh blocks from light to dark (white, yellow, orange, red) 
corresponds to the population density.
The thick and thin lines are 10 \% shortcuts 
and the other links in the LS network, respectively.
The size of a black node is proportional to the population in its
territory.}
\label{fig_visualize}
\end{figure}

\section{Efficiency and robustness} \label{sec3}
We investigate the topological structures of  LS, PR, and RS networks. 
Here, the inclusion of graphs and the extension with shortcuts are 
UDG $\supset$ LS 
$\stackrel{\rm extension}{\longrightarrow}$ PR or RS, 
and the positions of nodes are identical in these networks 
except for the removed parts.
As shown in Fig. \ref{fig_visualize}, 
an LS network is obtained from a UDG by removing some nodes 
in the (white) mountain or sea areas of low-population density, 
while the PR and RS networks have comparatively longer 
shortcuts between the areas of high-population density
and between random positions, respectively.
We remark that the LS network resembles a planar network without
crossing links. 
The large black nodes 
correspond to active nodes with more packet generations.
Surviving nodes naturally concentrate in dark (red) areas 
of large population similar to a router density map \cite{Yook02}
in the real world.
Here, 
the population in the territory of a removed node is reassigned 
to its neighboring nodes 
in the iterative process of network generation.
Remember that the initial distribution of 
population assigned to a node in the UDG 
or to the original mesh in $160 \times 160$
blocks decays exponentially (see Fig. \ref{fig_cum_pop}). 
This exponential-like distribution of population 
is consistent for LS networks after surviving some nodes. 
In the following, we will focus on the results obtained 
for the packet generation according to the population.
When the probability of packet generation at a node is set by 
$p_{a}$ and $p_{n}$ 
or by an exponential distribution $\exp( -\gamma \times nid)$, 
where $\gamma > 0$ is a parameter and 
$nid$ is an ID number: $1,2,\ldots,N_{0}$
randomly assigned to a node in the UDG with no relation to its 
geographical position, 
there are no changes qualitatively.

\begin{figure}[htb]
  \begin{minipage}[htb]{.47\hsize}
   \includegraphics[height=85mm,angle=-90]{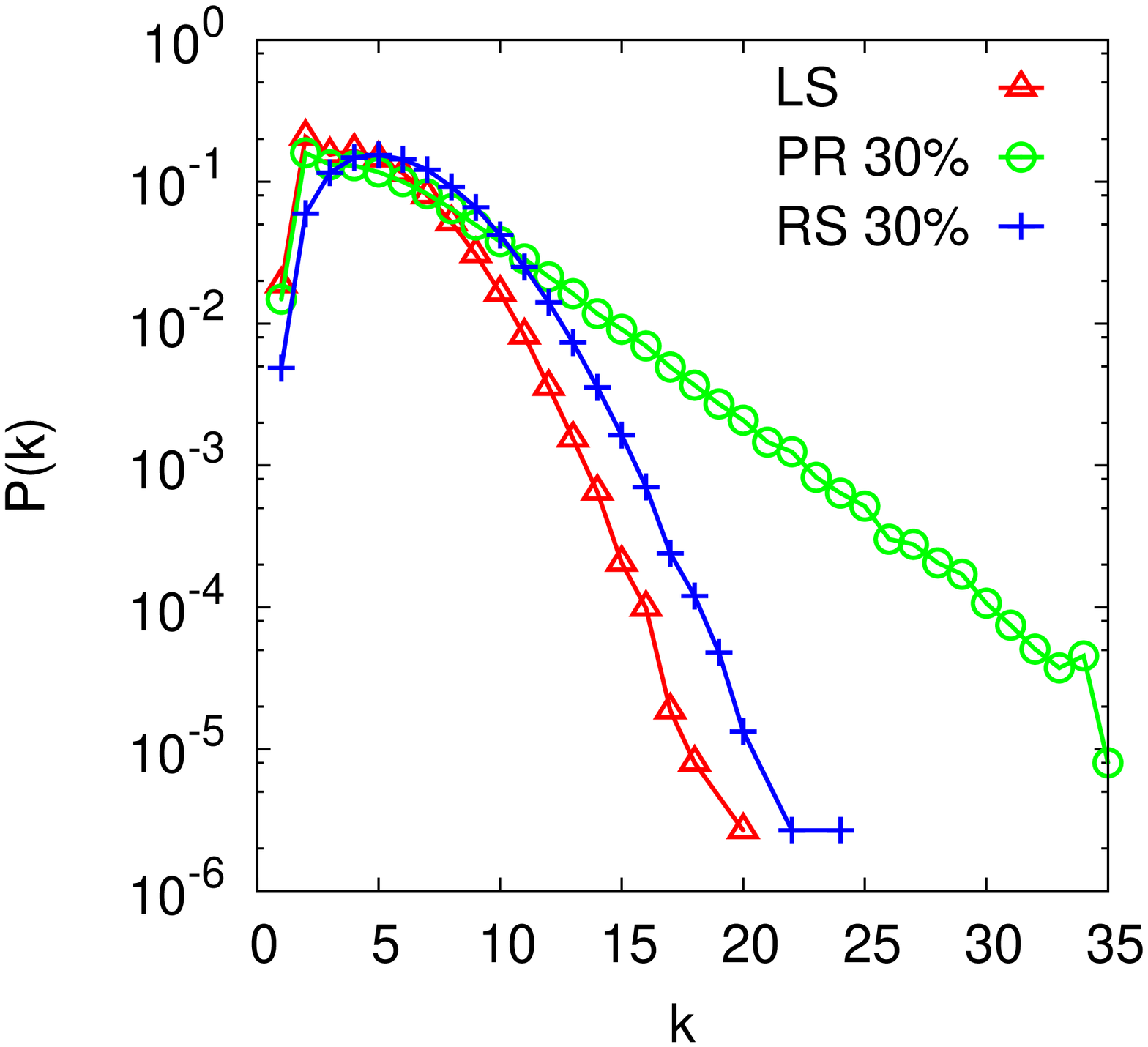} 
    \begin{center} (a) \end{center}
  \end{minipage}
  \hfill
  \begin{minipage}[htb]{.47\hsize}
   \includegraphics[height=85mm,angle=-90]{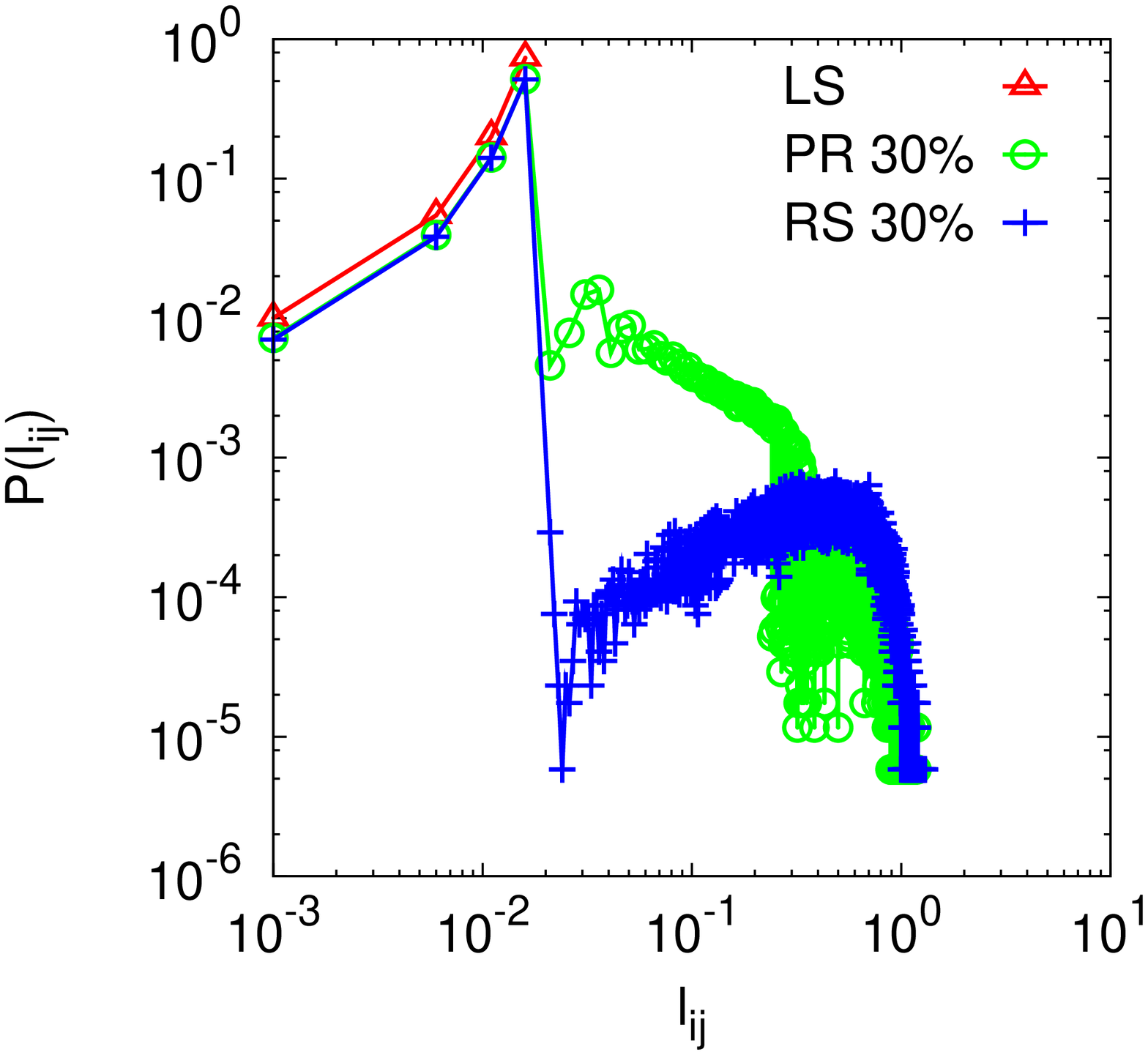} 
    \begin{center} (b) \end{center}
  \end{minipage}
\caption{(Color online) Distributions of (a) degree $k$
and (b) link length $l_{ij}$ in the networks 
generated from the initial size $N_{0} = 10000$. 
(a) Bounded small degrees are held even with shortcuts.
There exists no hub: the maximun degree is small as 
$k_{max} \sim O(\log N_{T})$, 
since the tail indicates an exponential decay.
(b) The right bottom parts marked by circle and plus signs denote 
longer shortcuts in PR and RS networks 
than the links in LS networks.
Each plot is obtained by the average of 50 realizations.}
\label{fig_pk_pl}
\end{figure}

\begin{table}[htb]
\begin{footnotesize}
\begin{center}
\begin{tabular}{c|cccccc} \hline\hline 
Case                       
 &  UDG &  LS  & PR10\% & RS10\% & PR30\% & RS30\% \\ \hline
Active-Normal &
   12.3 & 5.22 & 5.74 & 5.74 & 6.78 & 6.78 \\ 
Population & 
   12.3 & 4.56 & 5.01 & 5.01 & 5.92 & 5.92 \\ \hline\hline  
\end{tabular}
\end{center}

\vspace{2mm}
\begin{center}
\begin{tabular}{c|cccccc} \hline\hline   
Case
 &  UDG &  LS  & PR10\% & RS10\% & PR30\% & RS30\% \\ \hline
Active-Normal &
   0.013 & 0.016 & 0.035 & 0.06  & 0.069 & 0.129\\ 
Population & 
   0.013 & 0.016 & 0.033 & 0.058 & 0.065 & 0.123\\ \hline\hline 
\end{tabular}
\end{center}
\end{footnotesize}
\caption{Average degree $\langle k \rangle$ (top)
and link length
$\langle l_{ij} \rangle$ (bottom) in the networks 
with active-normal nodes and the corresponding results 
for the packet generation according to the population.
Each value is obtained by the average of $50$ realizations
generated from UDGs with the initial size $N_{0} = 10000$.}
\label{table_ave}
\end{table}

\vspace{2mm}
\begin{table}[htb]
\begin{footnotesize}
\begin{tabular}{c|ccccccc} \hline\hline 
$N_{0}$ & 
 100 & 200 & 500 & 1000 & 2000 & 5000 & 10000 \\ \hline
Active-Normal &
   55.26 & 130.66 & 379.3 & 823.92 & 1728.56 & 4517.74 & 9180.98 \\
Population & 
   61.94 & 127.0 & 331.26 & 688.94 & 1425.4  & 3679.98 & 7489.92 
\\ \hline\hline
\end{tabular}
\end{footnotesize}
\caption{The number $N_{T}$ of surviving nodes 
from the initial size $N_{0}$ of UDGs.
Note that the LS, PR, and RS networks have the same number of nodes.} 
\label{table_NT}
\end{table}

As typical measures for the communication efficiency, we investigate 
the degrees and the link lengths.
Because they are related to the costs for maintaining connections at a
node and for constructing a link by wireless beam or wire cable, 
smaller values are suitable for the efficiency. 
The costs are usually evaluated by the maximum (in the worst case)
and the average or the high-frequency (in a standard case) 
values over a network. 
Figures \ref{fig_pk_pl}(a)(b) show the distributions of degree 
and link length, and Table \ref{table_ave} shows their averages.
The increasing order of $\langle k \rangle$ is 
LS $<$ PR $\approx$ RS $<$ UDG, 
and that of $\langle l_{ij} \rangle$ is UDG $<$ LS $<$ PR $<$ RS. 
In particular, RS networks have on average twice as long links than PR
networks. 
It is a cause of the difference in $\langle l_{ij} \rangle$
that the length of a shortcut is affected 
by the paths of packets in PR networks
while it is determined only by the spatial distribution of surviving 
nodes which are embedded in RS networks. 
In addition, as shown in Fig. \ref{fig_pk_pl}(b), 
PR networks are more efficient in the sense of link length 
by far than RS networks with no restriction on the link length 
except the scale of the area.
On the whole, 
since the high-frequency parts of $P(k)$ and $P(l_{ij})$ consist of 
small values in spite of the effect of longer shortcuts 
added in the PR and RS networks, 
these networks are efficient with small degrees and short links.

\begin{figure}[htb]
  \begin{minipage}[htb]{.47\hsize}
   \includegraphics[height=85mm,angle=-90]{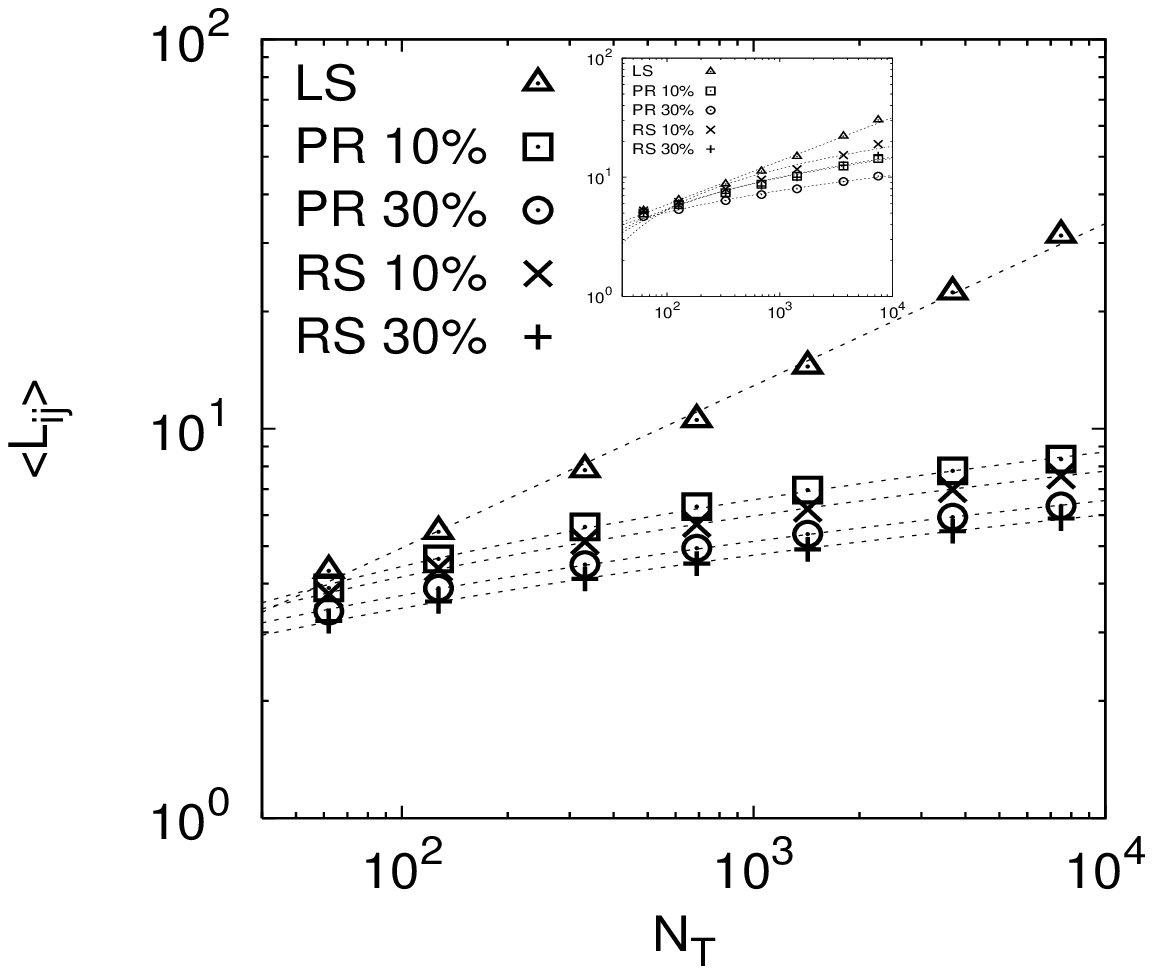} 
    \begin{center} (a)  \end{center}
  \end{minipage}
  \hfill
  \begin{minipage}[htb]{.47\hsize}
   \includegraphics[height=85mm,angle=-90]{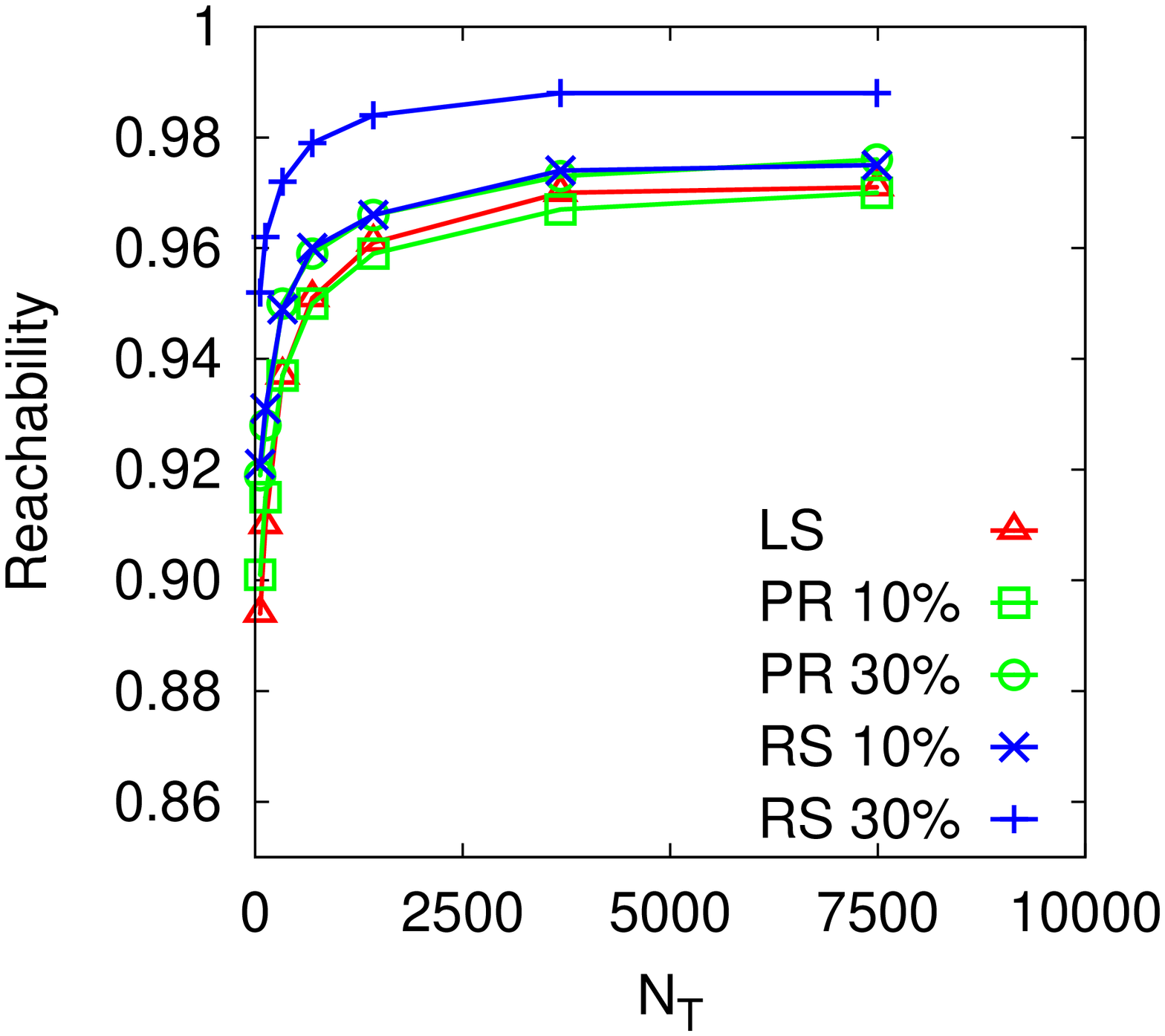} 
    \begin{center} (b)  \end{center}
  \end{minipage}
\caption{(Color online) Routing properties. 
(a) The scaling behaviors of the average number 
$\langle L_{ij} \rangle$ of hops on the shortest paths  
vs. the network size $N_{T}$. 
The estimated dashed lines are $O(N_{T}^{0.42})$
in LS networks and $O(\log N_{T})$ as the SW effect
in PR and RS networks with 10 $\%$ or 30 $\%$ shortcuts.
The inset shows the similar scaling behavior for the 
modified greedy routing. 
(b) The high reachability for the modified greedy routing. 
Each plot is obtained from the average of 50 realizations.}
\label{fig_smallworld}
\end{figure}

As another measure of the efficiency, 
the average number of hops $\langle L_{ij} \rangle$ on the
shortest paths (measured by the minimum number of hops between two
nodes) is important, 
because fewer hops are better for communicating through intermediate
nodes as quickly as possible. 
It is useful to investigate the scaling behavior of 
$\langle L_{ij} \rangle$ versus $N_{T}$ 
for checking the small-world (SW) effect 
\cite{Watts98,Albert99,Barthelemy99}.
As shown in Fig. \ref{fig_smallworld}(a),
the $O(\sqrt{N_{T}})$ scaling in LS networks 
is improved to $O(\log N_{T})$ characterized as the SW effect 
in PR and RS networks with only 10-30 \% of shortcuts. 
Surprisingly, this effect appears 
even in PR networks with more restricted shortcuts on the paths 
than in RS networks. 
Such change of scaling behaviors numerically occur 
within a transition of random-SW-regular phases \cite{Sen02}, 
when the addition of long distance connections on a $d$-dimensional
lattice is restricted by a probability $P(l) \sim l^{\ -\delta}$ 
for the distance $l$ with a control parameter $\delta \geq 0$. 
The related theoretical analysis is provided 
in \cite{Kleinberg00}; 
however it is still open for more general cases 
beyond lattice structures. 
On the other hand, 
$O(\sqrt{N_{T}})$ scaling usually appears 
in planar networks \cite{Barthelemy08}.
Table \ref{table_NT} shows the number $N_{T}$ of surviving nodes from 
the initial size $N_{0}$ of UDGs.
The above improvement is also obtained for other paths 
on the modified greedy routing; 
however, the improvement becomes smaller as shown in the inset.
Figure \ref{fig_smallworld}(b) shows the high reachability 
for the modified greedy routing in all of the networks. 
The reachability is defined by the fraction of packets successfully
arriving at the terminals (they are not stopped at dead-ends)
over all of the generated packets until $T$ steps
after constructing the networks.

\begin{figure}[htb]
  \begin{minipage}[htb]{.47\hsize}
   \includegraphics[height=85mm,angle=-90]{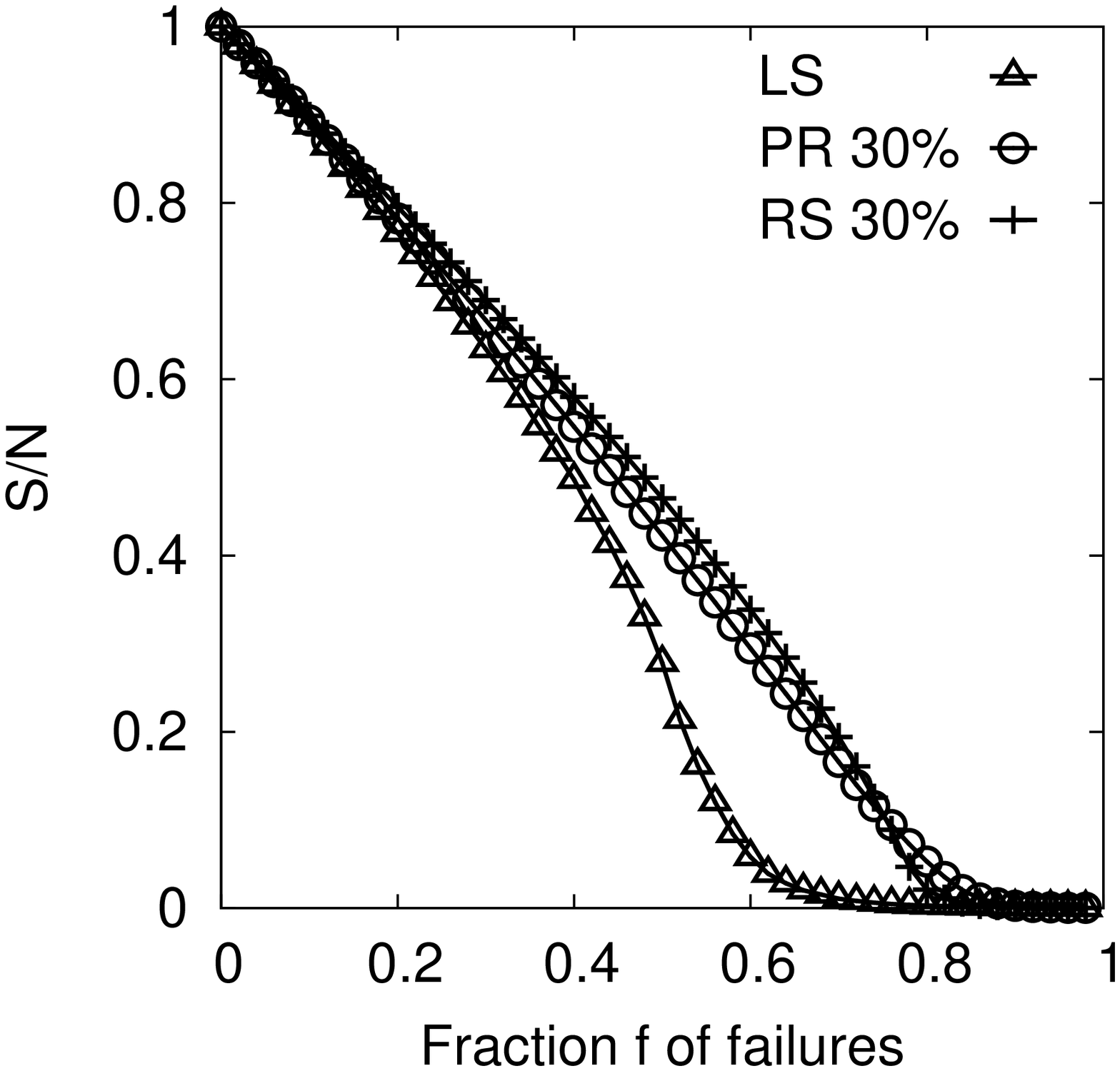} 
  \end{minipage}
  \hfill
  \begin{minipage}[htb]{.47\hsize}
   \includegraphics[height=85mm,angle=-90]{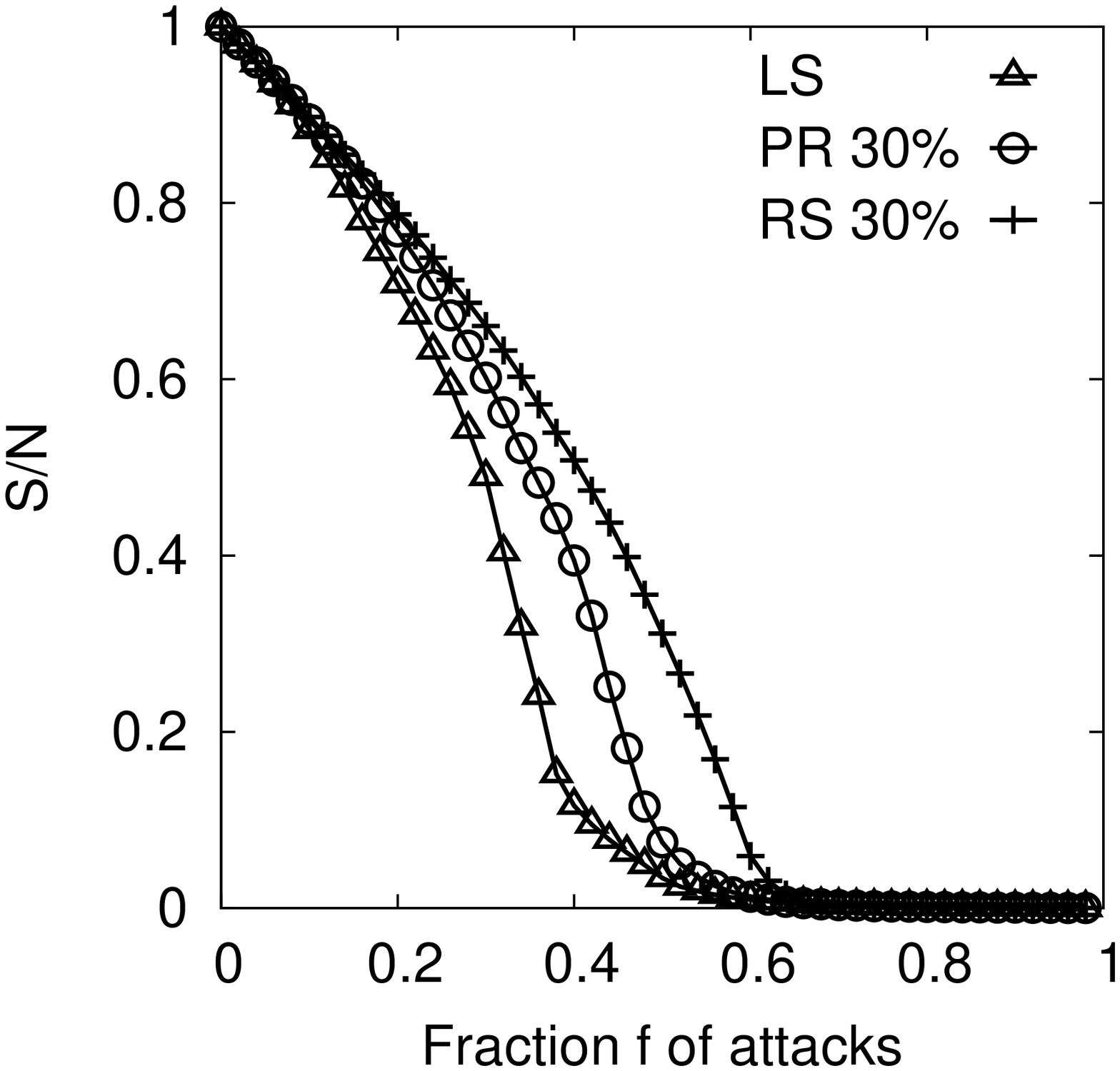} 
  \end{minipage}
  \hfill  \begin{minipage}[htb]{.47\hsize}
   \includegraphics[height=85mm,angle=-90]{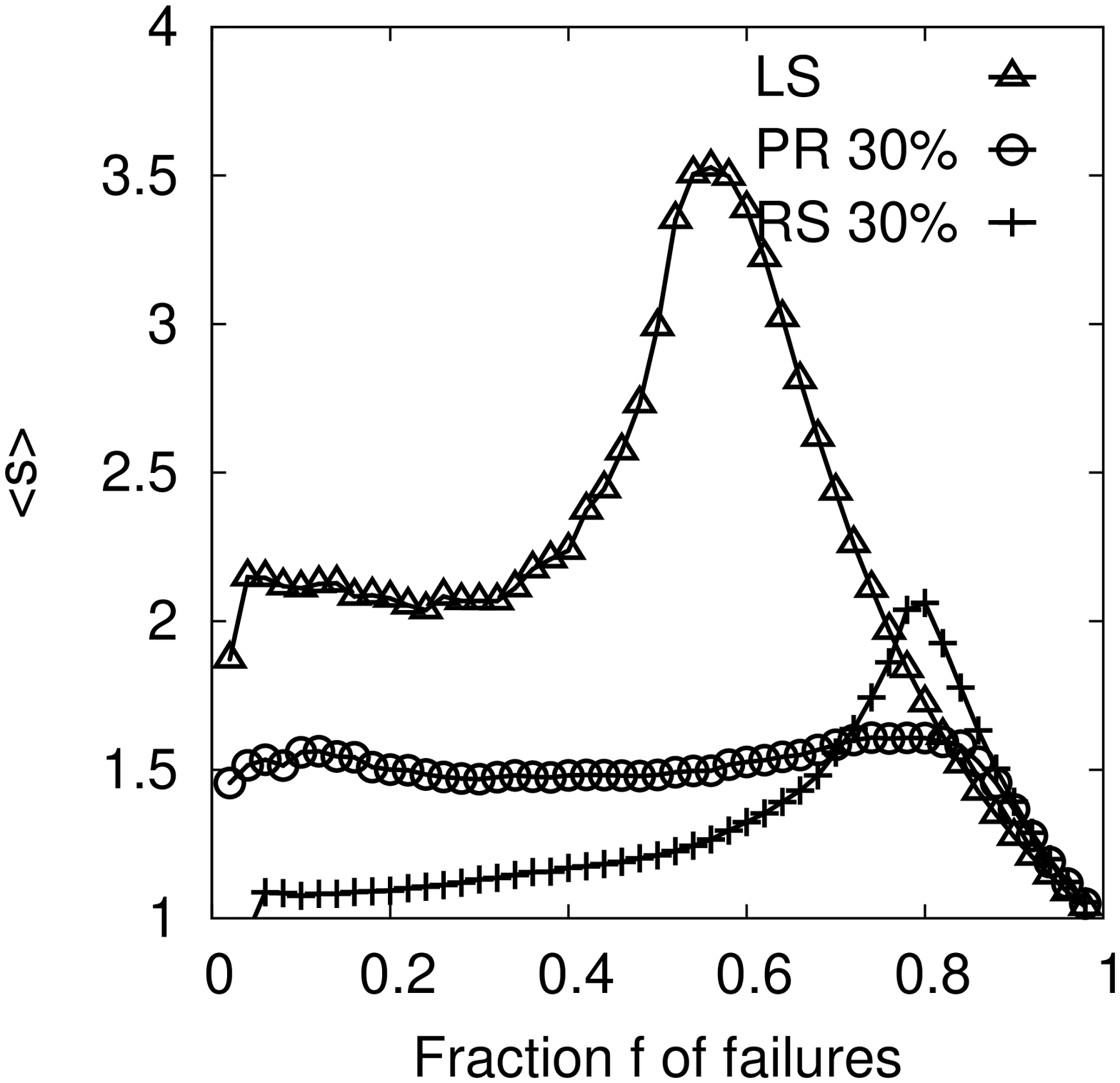} 
  \end{minipage}
  \hfill
  \begin{minipage}[htb]{.47\hsize}
   \includegraphics[height=85mm,angle=-90]{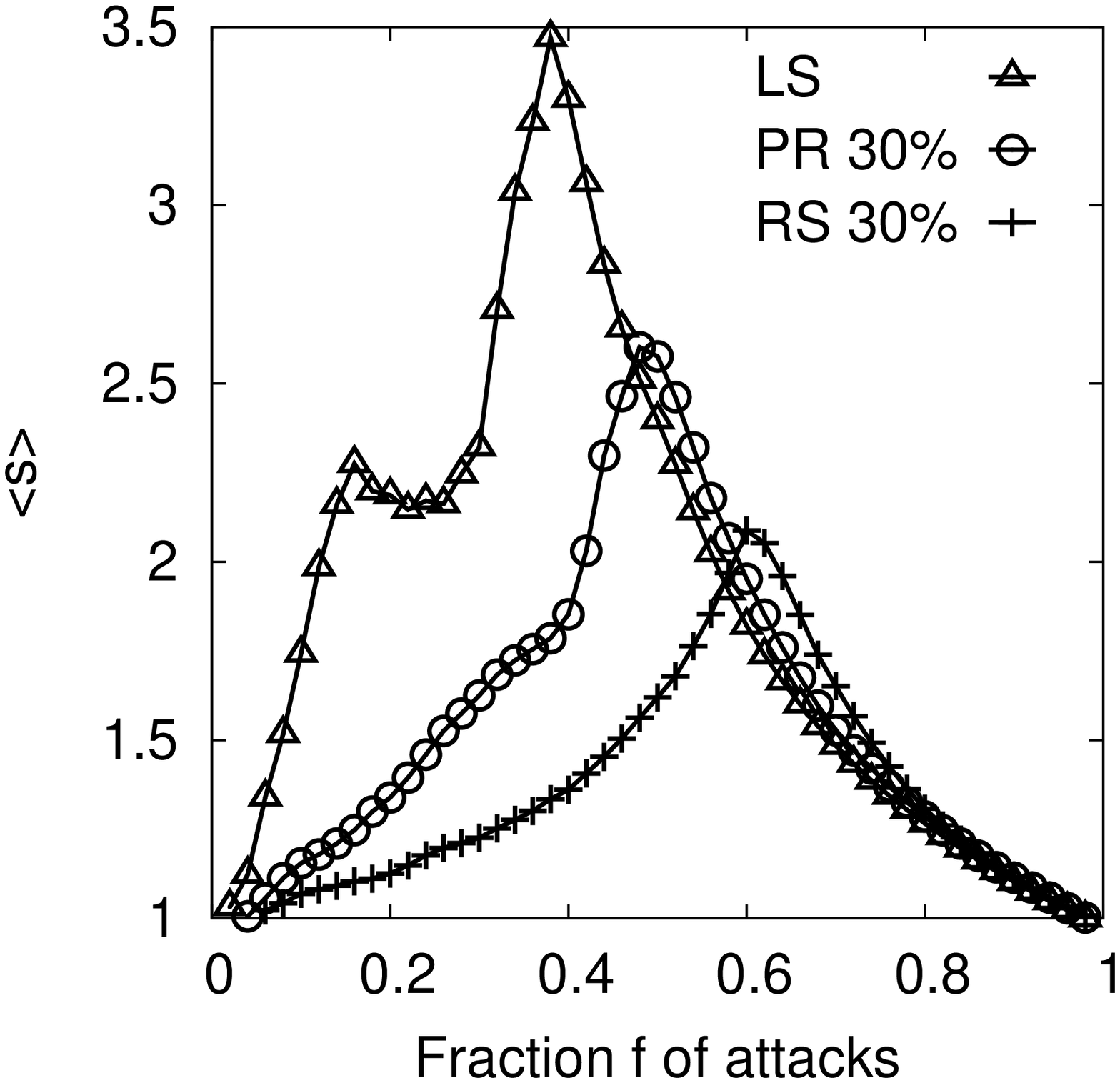} 
  \end{minipage}
\caption{Top: relative size $S/N$ of the GC against random failures (left)
and intentional attacks (right) in the networks 
generated from the initial size $N_{0} = 10000$.
Bottom: average size $\langle s \rangle$ of isolated clusters except the
GC.
Note that the GC is broken at the relative maximum of $\langle s \rangle$, 
whose fraction yields the critical value.
Each plot is obtained by the average of 50 realizations.}
\label{fig_robust}
\end{figure}

\begin{figure}[htb]
  \begin{center}
   \includegraphics[height=95mm,angle=-90]{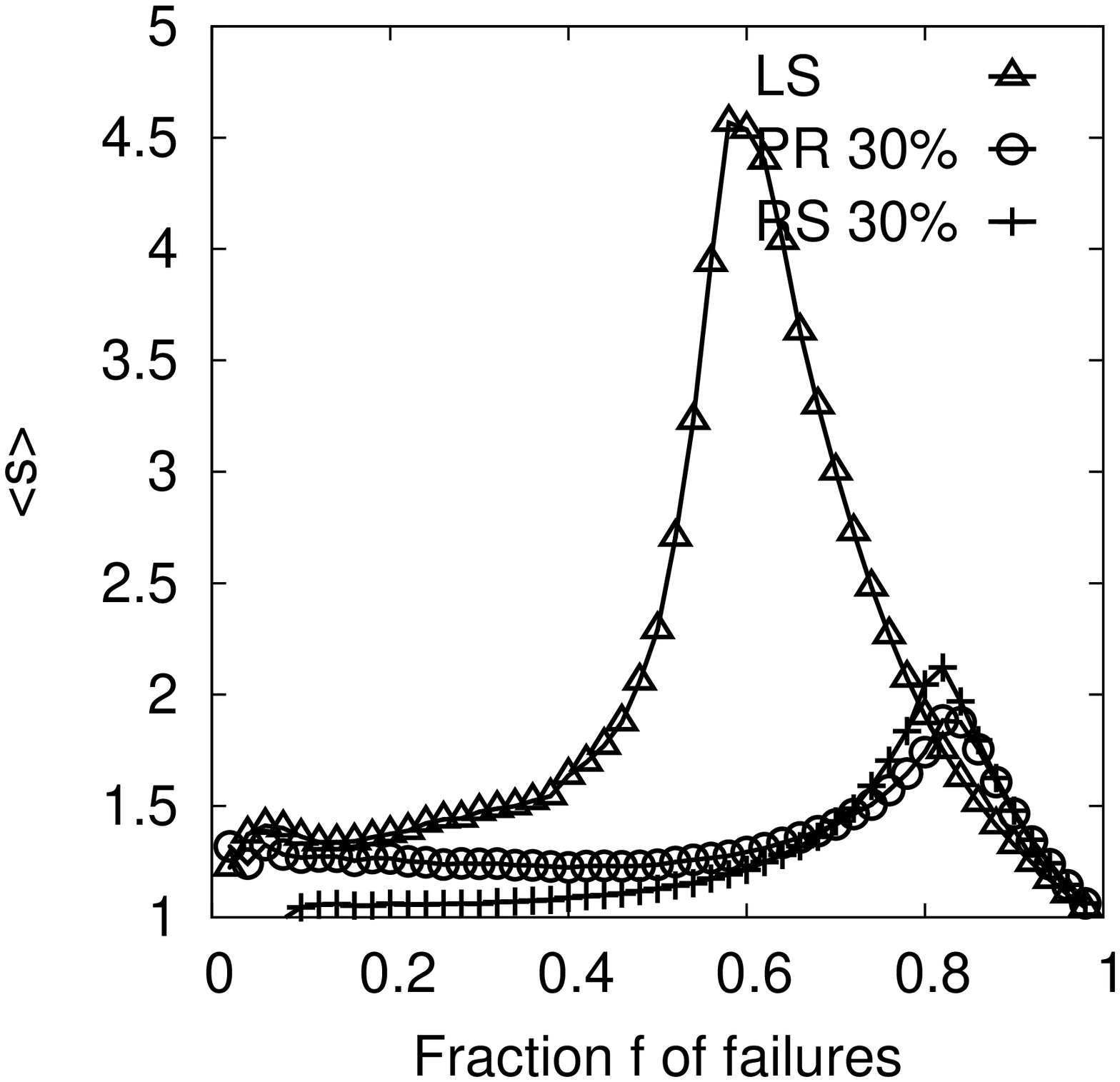} 
  \end{center}
\caption{Suggestion of percolation transition. 
The divergence peak appears 
when the probability of packet generation at a node is artificially set 
by $p_{a}$ and $p_{n}$ or by an exponential distribution with no
 relation to its geographical position.}
\label{fig_fs_exp}
\end{figure}

We confirm the effect of shortcuts on the robustness 
against random failures and intentional attacks, in which nodes are
removed at random 
and in decreasing order of degree before the removals, respectively.
Figure \ref{fig_robust} shows the relative size $S/N$ 
of a giant component (GC) at $N = N_{T}$
and the average size $\langle s \rangle$ of the isolated clusters
except the GC.
The critical fraction of failures for the breaking of GC
increases from 0.6 in LS networks 
to 0.8 in PR and RS networks with shortcuts.
The critical fraction of attacks also increases 
from 0.4 in LS networks to around 0.6 in PR and RS networks. 
In comparison within similar average degrees, 
this robustness is at the same level in the optimal bimodal networks
\cite{Tanizawa06}, 
and slightly stronger than that 
for Delaunay triangulations (DT) with $\langle k \rangle \approx 6$ 
\cite{Hayashi07}
and MSQ networks based on a self-similar tiling 
\cite{Hayashi09} with $\langle k \rangle = 4.54$.
Remember that \cite{Albert00} 
a GC is broken by attacking only a few percent of hubs
in SF networks. 
In addition, the interesting problem is whether the percolation
transition exists or not in our self-organized networks.
Especially, the critical point of $\langle s \rangle$
is indistinct for random failures 
in the PR networks marked by circles 
at the bottom left of Fig. \ref{fig_robust} in the cases of packet
generation according to the population (also for other data); 
however the divergence peak appears 
in the cases of an artificially defined 
probability distribution for packet generation
as shown in Fig. \ref{fig_fs_exp}. 
We will further investigate the reason why the difference occurs.

\section{Conclusion} \label{sec4}
In summary, 
we propose self-organized geographical networks by link survival and 
addition of shortcuts according to packet flows 
on decentralized routing in a wireless environment.
We show that 
the positions of surviving nodes naturally concentrate on the areas 
of high-population density, and that the links between them are short.
By adding only 10-30 \% shortcuts, 
the average number of hops on the routing paths is improved 
from $O(\sqrt{N_{T}})$ in LS networks 
to $O(\log N_{T})$ as the SW effect 
\cite{Watts98,Albert99,Barthelemy99}. 
Moreover, for both random failures and intentional attacks, 
the robustness of connectivity increases to reach the same 
level as in the optimal bimodal networks \cite{Tanizawa06}.
The uniformly random shortcuts are slightly better for the robustness
than the ones by the path reinforcement.
Probably, random shortcuts tend to be useful for a strong robustness
by bridging distributed local areas; 
however the optimal shortcuts are an open problem.
Thus, our self-organized networks with shortcuts
keep a high communication efficiency in SF networks,
but also overcome their vulnerability \cite{Albert00}. 
For other criteria, 
e.g. maximizing the end-to-end throughput \cite{Krause06}, 
it is an attractive issue 
to find the optimal network structure and the self-organization 
mechanism.
Various network formations by other weight update rules and routings
will be also further investigated 
for future communication infrastructures, 
smart grids, and urban planning.

\section*{Acknowledgment}
The authors would like to thank the anonymous reviewers for their
valuable comments to improve both insufficient descriptions 
and the readability of the manuscript.
This research is supported in part by a 
Grant-in-Aid  for Scientific Research in Japan, No. 21500072.


\end{document}